\documentclass[%
 reprint,
 amsmath,amssymb,
 aps,
]{revtex4-2}

\usepackage{graphicx}
\usepackage{dcolumn}
\usepackage{bm}
\usepackage[%
    colorlinks=true,
    pdfborder={0 0 0},
    linkcolor=red
]{hyperref}
\usepackage{braket}
\usepackage{amsmath}
\usepackage{multirow}

\begin{document}

\preprint{APS/123-QED}

\title{Triply Resonant Photonic Crystal Nanobeam Cavities for Unconditional Photon Blockade}

\author{Richard Dong$^1$}
\author{Abhinav Kala$^2$}
\author{Andrew Lingenfelter$^3$}
\author{Michael S. Polania Vivas$^4$}
\author{Matthew D. Stearns$^{5,6}$}
\author{Arka Majumdar$^{1,2}$}
\email{arka@uw.edu}
\affiliation{
  $^1$Department of Physics, University of Washington, Seattle, WA 98195
}
\affiliation{
  $^2$Department of Electrical and Computer Engineering, University of Washington, Seattle, WA 98195
}
\affiliation{$^3$Pritzker School of Molecular Engineering, University of Chicago, Chicago, IL 60637, USA.}
\affiliation{$^4$CUNY Graduate Center, 365 5th Avenue, New York, NY 10016}
\affiliation{$^5$Department of Physics, University of Massachusetts, Dartmouth, MA 02747}
\affiliation{$^6$Center for Scientific Computing and Data Science Research, University of Massachusetts, Dartmouth, MA 02747}

\date{\today}

\begin{abstract}
The development of many scalable quantum technologies requires single-photon nonlinearity, such as single-photon blockade, in solid-state systems. Recently, it has been shown that single-photon Fock states can, in principle, be unconditionally generated using arbitrarily small intrinsic optical nonlinearities in photonic cavities. We investigate the feasibility of such a scheme in achieving photon blockade in an on-chip silicon photonics platform. We show that a triply resonant nanobeam cavity pumped with three monochromatic lasers could achieve such functionalities with quality factors $\sim 10^7$ and effective mode volumes $\sim 10^{-2} \mu m^3$, for experimentally feasible incident powers. Using quantum optical simulations, we propose an experimental protocol to generate single photons under this scheme. The constraints on the cavity design and experimental conditions are thoroughly explored to determine feasible regimes of operation.
\end{abstract}

\maketitle


\section{Introduction}

Single-photon nonlinearity in the solid state is a critical requirement for developing scalable photonic quantum information processing units and quantum simulators, among other technologies \cite{kala2025opportunities}. Single-photon blockade is the most prominent example of single-photon nonlinearity \cite{birnbaum2005photon}. It involves creating effective repulsive interactions between two individual photons and has applications such as single-photon generation, quantum gates, and single-photon switches \cite{chang2014quantum}. The most common way to achieve single-photon blockade is by using a two-level quantum emitter strongly coupled to an optical cavity mode. The dynamics of this system could be described using the Jaynes-Cummings model \cite{shore1993jaynes}. The uncoupled cavity eigenstates for different photon numbers are evenly separated. However, in the strong-coupling regime, hybrid light$\mbox{--}$matter eigenstates are formed with uneven/anharmonic energy spacings. Then, if a single photon is coupled to the cavity from a classical pump resonant to induce a transition from the ground state to an excited state, the coupling of a second photon in the cavity is prevented since it is no longer resonant to any transition. Hence, effectively, a repulsive photon$\mbox{--}$photon interaction is created, which is essentially the single-photon blockade. 

In solid-state, epitaxial quantum dots coupled to photonic crystal cavities have been used for single-photon blockade \cite{faraon2008coherent}. One major limitation of this approach is the difficulty in the deterministic integration of emitters inside the cavities. This issue could be circumvented if one uses the intrinsic optical nonlinearity of the cavity material to achieve anharmonic energy levels. Such photon-blockade has been proposed for both $\chi^{(2)}$ and $\chi^{(3)}$ nonlinearities \cite{majumdar2013single,ferretti2012single}. However, experimental demonstrations have thus far remained out of reach due to the requirement of large quality factors and small mode volumes. 
The requirement of a large optical nonlinearity could be avoided by implementing what is called the unconventional photon blockade. Unconventional photon blockade involves inducing destructive interference between various pathways leading to a two-photon Fock state by employing appropriate pumping conditions \cite{bamba2011origin, snijders2018observation}. The limitations of these schemes are that the generated photon states are Gaussian squeezed states rather than the pure Fock states of the conventional blockade, and the single-photon generation probability is typically of the order of $10^{-2}$ \cite{flayac2017unconventional}. \\
More recently, a photon blockade scheme for Fock state generation was proposed that makes use of the $\chi^{(3)}$ optical nonlinearity of the cavity material \cite{lingenfelter1}. Importantly, this approach also works when the nonlinear interaction strength is much smaller than the cavity loss rate and requires pumping a single cavity mode with a one-photon and a two-photon pump. In this paper, we adapt this method for an integrated photonic platform and analyze the parameter space to reach the photon blockade regime with an experimentally feasible (albeit challenging) system.

Under rotating wave approximation, in the frame of one-photon pump frequency $\omega_1$, the Hamiltonian for this system has the following form \cite{lingenfelter1}
\begin{subequations}
\begin{equation}
\hat{H} = \hat{H}_0 + \hat{H}_{pump} + \hat{H}_{NL}
\end{equation}
\begin{equation}
\hat{H}_0 = \Delta \hat{a}^\dagger \hat{a}
\end{equation}
\begin{equation}
\hat{H}_{pump} =  \Lambda_1\hat{a}^\dagger  + \Lambda_2\hat{a}^\dagger \hat{a}^\dagger +h.c.
\end{equation}
\begin{equation}
\hat{H}_{NL} = U\hat{a}^\dagger\hat{a}^\dagger \hat{a}\hat{a}
\end{equation}
\label{Hamiltonian}
\end{subequations}
Where, $\hat{H}_0$ is the Hamiltonian in the absence of nonlinearity and external pumping, $\hat{H}_{pump}$ includes the pumping terms, $\hat{H}_{NL}$ accounts for the nonlinear $\chi^{(3)}$ interaction, and $h.c.$ stands for the Hermitian conjugate. $\Delta = \omega_1 - \omega_{c}+2U$ is the detuning of the pump to the cavity resonance where $\omega_{c}$ is the frequency of the cavity mode. $\hat{a}$ ($\hat{a}^{\dagger}$) is the destruction (creation) operator, $\Lambda_1$ is the single-photon-pump amplitude, and $\Lambda_2$ is the two-photon-pump amplitude. The mechanism to achieve a two-photon drive is discussed in the next section. By choosing the appropriate parameter values, one can create an effective $n$-level system, where $n$ is a positive integer, in the displaced frame with a displacement parameter $\alpha$ which is determined by the nonlinear interaction strength, as shown in equation \ref{Lambda_NL} below. The target Hamiltonian and the parameters have the following form \cite{lingenfelter1}

\begin{equation}
\hat{H}_{block} = \Lambda_{NL}\hat{a}^\dagger(\hat{a}^\dagger \hat{a}-n) + h.c.
\label{H_block}
\end{equation}
where 
\begin{equation}
 \Lambda_{NL} = 2U\alpha
 \label{Lambda_NL}
\end{equation}

The parameter choices to get this Hamiltonian form are
\begin{subequations}
\begin{equation}
\Lambda_1 =  \Lambda_{NL}\left[\frac{\Lambda_{NL}^2}{2U^2}-n + \frac{i\kappa}{4U}\right]   
\end{equation}
\begin{equation}
\Lambda_2 =  -\Lambda_{NL}^2/4U     
\end{equation}
\begin{equation}
\Delta = -|\Lambda_{NL}|^2/U    
\end{equation}
\label{Parameters}
\end{subequations}
where $\kappa$ is the cavity mode decay rate. It can be seen from equation \ref{H_block} that the only non-zero matrix elements of $H_{block}$ are of the form  $|m+1\rangle\langle m| + h.c.$. Among these, for perfect tuning, i.e. \textendash{} when $n$ is an integer, $|n+1\rangle\langle n|$ term vanishes. Therefore, for $n=1$, this is an effective two-level system and could lead to single-photon blockade. This theoretical proposal is agnostic to the type of cavity and therefore requires further investigation for the specific material platform and corresponding spectral region, depending upon the chosen quantum system. In this work, we study the feasibility of this protocol on a silicon photonics platform operating at 1550 $nm$ wavelength. We study various requirements and estimate constraints for the cavity parameters and experimental implementation. 

\subsection{Experimental Proposal}

For the scheme described above, we consider a low-mode-volume silicon nanobeam cavity with three equally separated modes at angular frequencies $\omega_{1c}$, $\omega_{2c}$, and $\omega_{3c}$, which we call mode 1, mode 2, and mode 3, respectively. Also, as shown in Fig. \ref{fig-Experimental schematic}, $\omega_{3c} = \omega_{1c} + \delta = \omega_{2c} +2\delta$. Triply resonant cavities have been previously proposed in the literature and could also be created using inverse design \cite{buckley2014multimode}. We use $\omega_{1c} = 1.215\times10^{15}\mathrm{\,Hz}$, which corresponds to a free-space wavelength of $1550 \mathrm{\, nm}$. The cavity modes are chosen to facilitate the dual-pump degenerate four-wave mixing $\omega_{2c} + \omega_{3c} = 2\omega_{1c}$ \cite{he2023conditions}. This process essentially creates two photons at frequency $\omega_{1c}$ by annihilating one photon each at frequencies $\omega_{2c}$ and $\omega_{3c}$. This is how we create the two-photon drive. The Hamiltonian term for this process ($\hat{H_2}$) can be written as 
\begin{equation}
\hat{H_2} = \beta \hat{a_2}\hat{a_3}\hat{a_1}^{\dagger}\hat{a_1}^{\dagger} + h.c.   
\label{H_2ph}
\end{equation}
where the coupling constant $\beta$ can be written as an integral, given by equation \ref{beta_eq} as
\begin{equation}
\beta = \frac{3\epsilon_0}{8}\sqrt{\frac{\omega_{1c}^2 \omega_{2c} \omega_{3c}}{ V_1 ^2 V_2 V_3 \bar{\epsilon}_{1}^2 \bar{\epsilon}_{2} \bar{\epsilon}_{3}}}\int d^3\vec{r} \chi^{(3)}(\vec{r}) \phi_1^{*2}(\vec{r})\phi_2(\vec{r})\phi_3(\vec{r})
\label{beta}    
\end{equation}
Here $\bar{\epsilon}_{i}$, $\phi_i$, and $V_i$ are the reference permittivity (permittivity at the field maxima), the mode field profile, and the mode volume for $i^{th}$ mode, respectively, such that electric field for the $i$th mode is $\hat{E}_i(\vec{r}) = \sqrt{\frac{\omega_{ic}}{2\bar{\epsilon}_i V_i}}\phi_i (\vec{r}) \hat{a}_i + h.c. $ and $V_i = (\int d^3 \vec{r'} \frac{\epsilon_i(\vec{r'})}{\bar{\epsilon_i}}|\phi_i (\vec{r'})|^2) $. $\epsilon_i(\vec{r'})$ is the permittivity for mode $i$ at location $\vec{r'}$. The expression for $\beta$ is derived in the Appendix \ref{beta}. And, $\chi^{(3)}(\vec{r})$ is the position-dependent third-order nonlinear susceptibility. We further assume that the pumps at mode 2 and mode 3 are strong classical pumps and can be replaced with corresponding classical mode fields $a_2$ and $a_3$, respectively
\begin{equation}
\hat{H}_{2ph} = \beta a_2 a_3\hat{a_1}^{\dagger}\hat{a_1}^{\dagger} + h.c.    
\end{equation}
Comparing with the expression for $\Lambda_2$ in equation \ref{Hamiltonian}, we find that
\begin{equation}
\Lambda_2 = \beta a_2a_3   
\label{Lambda_2}
\end{equation}
A schematic for this experiment is provided in Figure \ref{fig-Experimental schematic}.

Using the cavity and material parameters, we can estimate the nonlinearity amplitude $U$ using the following relation \cite{ferretti2012single}
\begin{equation}
U = \frac{3(\hbar\omega_{1c})^2 \chi^{(3)}}{4\epsilon_0V_{eff}\epsilon_r ^2 }
\end{equation}

Here, $V_{eff} = \int |\vec{\phi}_1(\vec{r})|^4dV$ is the effective mode volume of mode 1, where $\vec{\phi}_1(\vec{r})$ is the normalized mode-field profile and $\epsilon_r$ is the relative permittivity of the silicon.

While quality factors and effective mode volumes for cavity modes are treated as variables for some of the calculations in this work, we also use one standard set of parameters. For mode 1, we assume a realistic effective mode volume of $0.01\mathrm{\, \mu m^3} $ and a quality factor of $10^7$, corresponding to a decay rate $\kappa = 193.4 \mathrm{\, MHz}$. The real part of the average nonlinear susceptibility of silicon is $\chi^{(3)} = 0.45\times10^{-18} \mathrm{\, m^2/V^2}$. Using the standard parameters, we find that $U=4.4 \mathrm{\, MHz}$, which corresponds to $U/\kappa \approx 0.023$.

The protocol, as introduced in the previous work, can be divided into three steps, as shown in Fig. \ref{fig-Experimental schematic} \cite{lingenfelter1}. The initialization step applies a displacement operation on the cavity field. After that, the system evolves under the Hamiltonian $\hat{H}_{block}$. At the end of this step, the cavity is in the one-photon Fock state in the displaced frame ($|\alpha\rangle_1$) with non-zero probability, which is converted to a one-photon Fock state in the lab frame in the final displacement step. 

\subsection{Methods}%
\begin{figure*}
\includegraphics[width=\textwidth]{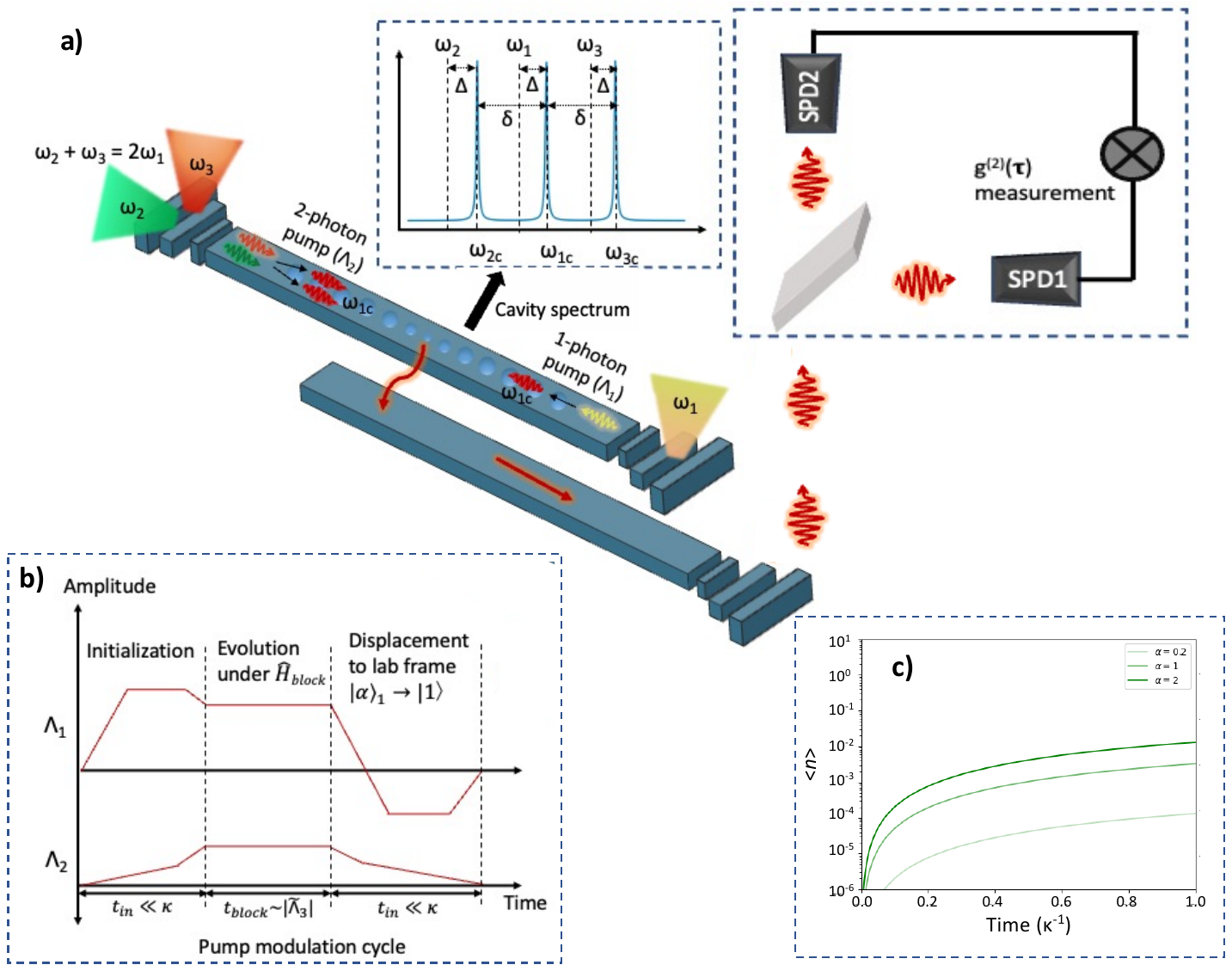}
\caption{\label{fig-Experimental schematic} a) Schematic of the Experiment $\mbox{--}$ The triply resonant cavity with $\chi^{(3)}$ nonlinearity has three modes at $\omega_{1c}$, $\omega_{2c}$, and $\omega_{3c}$. The one-photon pumping is implemented using a laser at $\omega_1$. For two-photon pumping, two separate lasers can be used at $\omega_{2}$ and $\omega_{3}$ such that $\omega_{2}+\omega_{3}  = 2\omega_{1}$. b) Pumping scheme for one and two-photon pumping $\mbox{--}$ the initialization phase lasts a short time $\tau \ll \kappa^{-1}$, where $\kappa$ is the cavity mode decay rate. It consists of three steps for the single-photon drive ($\Lambda_1$) and two steps for the two-photon drive ($\Lambda_2$), which are visible as linear steps with different slopes in the plot. The rationale behind such initialization is discussed in \cite{lingenfelter1}. $\Lambda_1$ is initially ramped up to a large magnitude, followed by a duration of constant value. $\Lambda_2$, on the other hand, is relatively slowly ramped up in the first step. In the next step, $\Lambda_1$ and $\Lambda_2$ are linearly ramped down and up, respectively, such that the desired displacement as well as the Hamiltonian $\hat{H}_{block}$ is achieved at the same time. During the phase of evolution under $\hat{H}_{block}$, the drives are held at a steady level for a time $t$ on the order of $1/\Lambda_{NL}$. The final phase, displacement to the lab frame, is a reversal of the initialization phase. At the end of the pumping sequence, the photon is collected via a side-coupled waveguide. The emitted photons at $\omega_1$ will be sent to a Hanbury Brown--Twiss interferometer to characterize the second-order time correlation function ($g^{(2)}(\tau)$) of the single photon emission. c) Time evolution of the cavity photon number $\mbox{--}$ The time-dependent average cavity photon number ($\langle n \rangle$) during the evolution under the blockade Hamiltonian $\hat{H}_{block}$ is plotted for multiple values of $\alpha$. The plot here is for the ideal case with no errors in any experimental parameter.}
\end{figure*}

We simulated the system using the Quantum Toolbox in Python (QuTiP) \cite{qutip5}. The power requirements for the experimental regime are calculated using cavity input--output theory. For simulations, the system is modeled as a single-mode cavity (with the mode at $\omega_{1c}$) evolving under the master equation

\begin{equation}
\frac{d\hat{\rho}}{dt} = - i[\hat{H}, \hat\rho] + \kappa \mathcal{D}[\hat{a}]\hat\rho
\label{Master_eq}
\end{equation}

with single-photon loss modeled as the dissipator $\mathcal{D}[\hat{a}]\hat\rho = \hat{a}\hat\rho\hat{a}^{\dagger} - \frac{1}{2} \{\hat{a}^\dagger \hat{a}, \hat\rho\}$ with rate $\kappa$. In the simulations, we use $\alpha = 2$ due to computational limitations. As shown in Fig. \ref{fig-Experimental schematic}, for the standard experimental parameters, this choice of $\alpha$ corresponds to a single-photon generation probability of 1.3\%.

To find the ideal time-dependent drive amplitudes for initialization, we used gradient descent. There are multiple ways to optimize the system, and we optimized over the ``nearness" of the initialized state to the desired coherent state. We took the difference between the lower-moments of the initialized state compared to their expected values, i.e. $\langle \hat{a}\rangle - \alpha$, $\langle \hat{a}^2 \rangle - \alpha^2$, etc. The loss function is the sum of the weighted amplitude of the initial four moments $\sum_{i=1}^4c_i|\delta_i|$, where $\delta_1 = \langle \hat{a}\rangle - \alpha$, $\delta_2 = \langle\hat{a}^\dagger \hat{a}\rangle - |\alpha|^2 $, $\delta_3 = \langle \hat{a}^2 \rangle - \alpha^2 $, and $\delta_4 = \langle \hat{a}^3 \rangle - \alpha^3$, with weights $c_1 = 10$ and $c_2 = c_3 = c_4 = 1$. To second-order, this loss function gave reasonably optimized values for $\Lambda_1$ and $\Lambda_2$. The difference between the loss model and the actual $g^{(2)}(0)$ is discussed in Section II(B).

Furthermore, only linear time dependence was assumed for the three sub-steps of the initialization step,
shown in Fig. \ref{fig-Experimental schematic}(b). However, this loss function is not convex over the parameter space, and in fact the system itself possesses multiple local minima for the zero-delay second-order time-correlation function $g^{(2)}(0)$, which is discussed in more detail in section II(B). To bypass this, we choose as a starting point an initial $\Lambda_1$ parameter that would initialize the desired state for a linear ($U=0$) cavity. Assuming weak nonlinearity, the same $\Lambda_1$ parameter will initialize a state close to the desired state when we include $U$. From here, we can run gradient descent as the loss function is locally convex near the optimal minimum. For estimation of errors in implementation, we ran parameter sweeps to determine the effects of initialization time, value of $\alpha$, and inaccurate initialization. The time dependence of the two pumps is shown in Fig. \ref{fig-Experimental schematic}, where we can see that the displacement should be complete when both drives reach the values provided in equation \ref{Parameters}.

\section{Experimental Feasibility Study}
In this section, we investigate the requirements for experimental feasibility of this protocol. For that, we first estimate the required pump powers for different cavity parameters. Furthermore, we also estimate the necessary timescale for pump power modulation and the effect of experimental errors on the blockade.

\subsection{Power Requirements}
First, we estimate the power required during evolution in the displaced frame under the Hamiltonian $\hat{H}_{block}$. Using the standard input$\mbox{--}$output relations, we can express the input power for the one-photon pump ($P_1$) by the following expression using equation (\ref{power_1}),  

\begin{equation}
P_1 = \frac{\hbar\omega|\Lambda_1|^2}{\kappa} 
\end{equation}

\begin{figure}
\includegraphics[width=\linewidth]{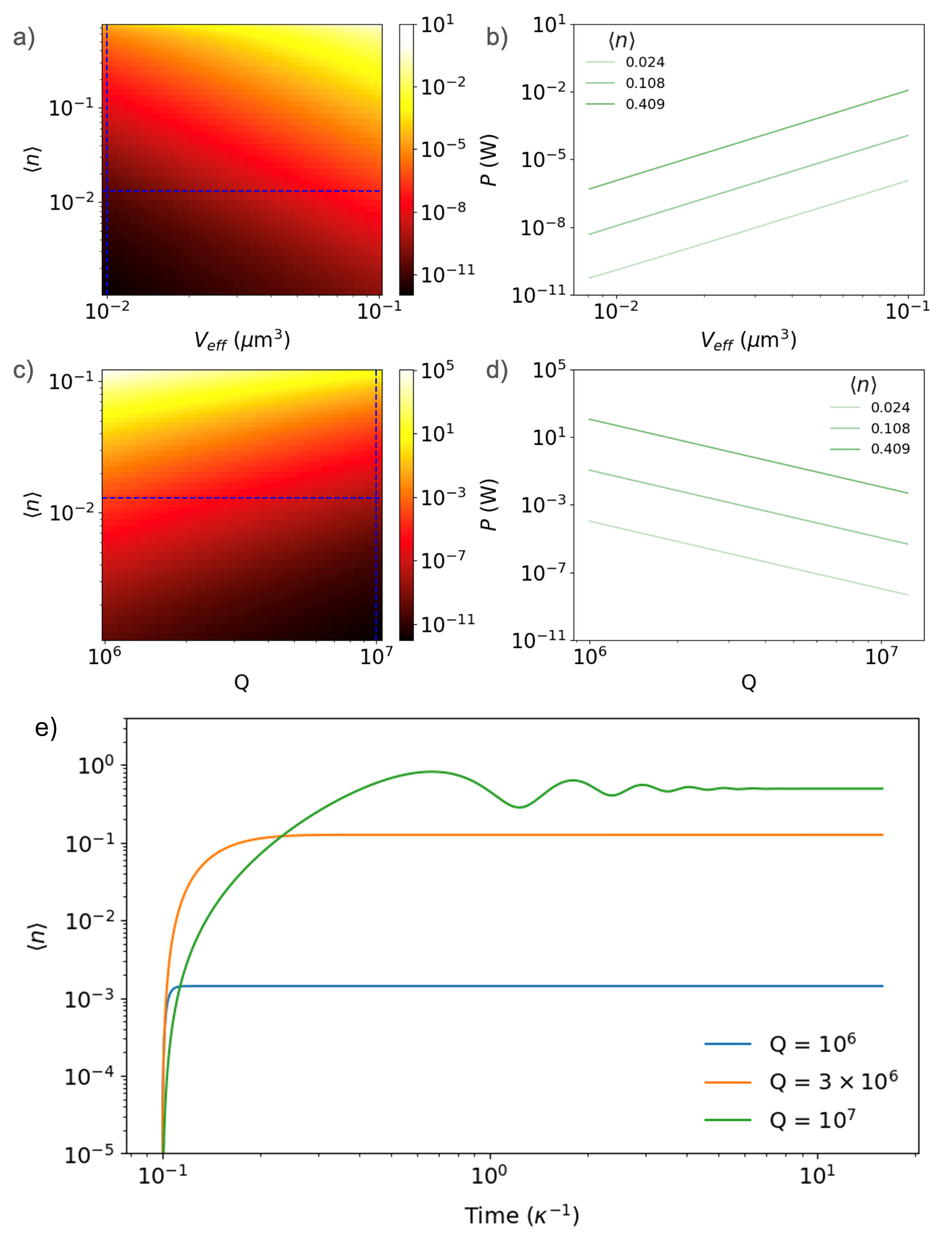}
\caption{\label{fig:V vs n vs Lambda_1} a) Power dependence of the single-photon pump as a function of the average photon number and mode volume of the cavity. Here, $Q = 10^5$, $\omega_c = 1.21 \times 10^{15} s^{-1}$. b) Power dependence with the mode volume of the cavity for a desired photon number. This is equivalent to a horizontal line in a). c) Power dependence of the single-photon pump as a function of the average photon number and quality factor of the cavity. Here, $V = 0.01\mathrm{\,\mu m^3}$, $\omega_c = 1.21 \times 10^{15}$ $\mathrm{Hz}$. d) Same as b), but as a function of the quality factor. The dotted lines represent the parameters assumed for other simulations. e) The time-dependent expected photon number in the displaced frame for three different cavity quality factors and the identical input power of 10 ${\,mW}$.}
\end{figure}

Furthermore, using the expressions for $\Lambda_1$ from equation (\ref{Parameters}), we can estimate the power requirement as a function of quality ($Q$)-factor and mode volume $V_{eff}$ for mode 1. In Fig. \ref{fig:V vs n vs Lambda_1}(a), we have plotted the power dependence of $P_1$ as a function of the cavity mode volume and average photon number ($\langle n\rangle$) for a fixed $Q$-factor $10^7$, assuming an ideal case with no errors and perfect displacement in the initial and final step. $\langle n\rangle$ also represents the probability of generating a single photon. Fig. \ref{fig:V vs n vs Lambda_1}(b) demonstrates the mode volume dependence of $P_1$ for various fixed $\langle n\rangle$ values, which are extracted from Fig. \ref{fig:V vs n vs Lambda_1}(a). It can be seen from Fig. \ref{fig:V vs n vs Lambda_1}(b) that $P_1$ roughly scales as $V_{eff}^4$. 
Similarly, in Fig. \ref{fig:V vs n vs Lambda_1}(c) the power requirement is plotted as a function of $\langle n\rangle$ and the $Q$-factor for a fixed $V_{eff} = 0.01$ $\mu m^3$ along with certain constant $\langle n\rangle$ contours in Fig. \ref{fig:V vs n vs Lambda_1}(d). Furthermore, for a fixed $\langle n\rangle$, $P_1$ roughly scales as $Q^{-4}$. 

To avoid excess heating in silicon photonics coming from two-photon absorption, it would be preferred to keep the power limited to around $10 \mathrm{\, mW}$. From equations \ref{H_block} and \ref{Master_eq}, we can see that the dynamics of the system for $n=1$ is completely determined by the strength of the nonlinear drive $\Lambda_{NL}$, cavity decay rate $\kappa$, and nonlinearity amplitude $U$. Since $\Lambda_{NL}$ is proportional to both $U$ and $\alpha$, in principle, a smaller value of $U$ could be compensated by using a larger displacement, which is implicit in Fig. \ref{fig:V vs n vs Lambda_1} as well. However, doing so can lead to the requirement of experimentally unfeasible $P_1$. It can be seen more explicitly with the help of Fig. \ref{fig:V vs n vs Lambda_1} (e), where we have plotted the time-dependent photon-number $\langle n \rangle$ in the displaced frame for three different values of $Q$, while keeping the $P_1 = 10 \mathrm{\, mW}$. The effective mode-volume is kept fixed, and therefore $U$ is also fixed with a value $4.4 \mathrm{\, MHz}$. We can see that $\langle n \rangle$ goes down significantly with decreasing $Q$-factor. The maximum values of $\langle n \rangle$ are approximately $0.001$, $0.13$, and $0.81$ for $Q = 10^6$, $3 \times 10^6$, and $10^7$, respectively. In fact, to get $\langle n \rangle = 0.5$ with $Q = 3 \times 10^6$, the required power is more than 25$\mathrm{\, W}$. This rapid scaling of $P_1$ can be seen from equation \ref{Parameters}(b) as well. On the right-hand side, for large $\alpha$, the first term dominates, and we get $\Lambda_1 \propto \alpha^3$ and hence $P_1 \propto \alpha^6$. These estimations also justify the choice of $Q=10^7$ and $U = 4.4 \mathrm{\, MHz}$ as the standard parameters.   

We also estimate the power required for the two-photon driving lasers. For this, we assume that both mode 2 and mode 3 have the same loss rates ($\kappa_2$) and opposite detunings ($\pm\Delta_2$) and use the coupled mode equations for a triply resonant cavity \cite{ramirez2011degenerate}. For this estimation, we use classical pumps. For cavity mode field $a_i$ and corresponding input field operator $\hat{a}_{i,in}$, the equations are 

\begin{subequations}
\begin{equation}
\frac{da_2}{dt} = -i\Delta_2a_2 -\frac{\kappa_2}{2}a_2+ \beta \langle \hat{a}_1^2\rangle a_3^* + \sqrt{\kappa_2}a_{2,in}
\end{equation}
\begin{equation}
\frac{da_3}{dt} = i\Delta_2a_3 -\frac{\kappa_2}{2}a_3 + \beta \langle\hat{a}_1^2\rangle a_2^* + \sqrt{\kappa_2}a_{3,in}
\end{equation}
\end{subequations}

where $\beta$ is defined in equation \ref{beta}. While the exact value of $\beta$ depends on the particular nanobeam design, values significantly larger than $0.01U$ are achievable \cite{lin2014high}. To avoid underestimation of pump powers, we assume it to have a relatively small value of $0.01 U$. Since we want to do an order-of-magnitude estimation, we have ignored the cross-phase modulation and self-phase modulation terms in the equations above. A more precise estimation should include those nonlinear terms as well. Furthermore, we assume that the inputs in modes 2 and 3 are strong enough to be approximated as classical fields. Then, in the steady state, the mode fields are
\begin{equation}
a_i = \frac{\sqrt{\kappa_2}a_{i,in}}{\frac{\kappa_2}{2}-i\Delta_i}
\label{a_i}
\end{equation}
where $i \in \{2,3\}$. From cavity input$\mbox{--}$output relations, we know that
\begin{equation}
|a_{i,in}|^2 = \frac{P_{i,in}}{\hbar\omega_i}
\label{P_i}
\end{equation}
As we can see from the pumping scheme in Fig. 1, the steady-state estimation should work as an upper bound for the $P_2$ and $P_3$ during the entire pumping cycle. Using expression for $\Lambda_2$ in equation (\ref{Lambda_2}) along with equations (\ref{a_i}) and (\ref{P_i}), we find that
\begin{equation}
P_2(\Delta) = \frac{1}{\beta}|\Lambda_2|\frac{(\kappa/2)^2  + (\Delta_2)^2}{2\kappa}\times\hbar\sqrt{\omega_1\omega_2}
\end{equation}

Here we have assumed the decay rates for mode 2 and 3 to be the same as the decay rate of mode 1, $\kappa$. For the order-of-magnitude estimation, we assume that $P_2 = P_3$. From equations \ref{Parameters} and \ref{Lambda_NL}, we can estimate the $P_2$ for any $\Lambda_{NL}$. In particular, for the previously discussed case of $P_1 = 10 \mathrm{\, mW}$ and $Q=10^7$, we find that $P_2 \approx 2 \mathrm{\, \mu W} $. This is the upper limit of two-photon pump powers since the magnitude of $\Lambda_2$ is maximum during this step of the scheme. This number is significantly smaller than the power required for the single-photon drive. Hence, from the standpoint of required pump power, two-photon drive is not the limiting factor in the implementation of this scheme.

The initialization step for the one-photon drive also consists of three parts: a ramping phase, a steady phase, and another ramping phase to match the drive for the next step of the protocol. An estimate for the steady-phase drive amplitude and power could be calculated under the assumptions of short ramp-up and ramp-down times in comparison to steady-phase time, in addition to initialization time being much shorter than $\kappa$. Then for a linear cavity, using equation (\ref{eq:final}) we find that
\begin{equation}
\Lambda_{1,in} = i\alpha/\tau
\label{lambda_1 linear}
\end{equation}

For parameters $\alpha=60.3$ and $\tau = 0.01\kappa^{-1} = 0.82 \mathrm{\, ns}$, using equation (\ref{power_1}), the initialization power required is $3.57 \mathrm{\,\mu W}$, which significntly smaller than the power required after the displacement, as calculated earlier in this section. As we will see in the next section, this choice of $\tau$ does not lead to significant errors in the protocol. Furthermore, this $\Lambda_{1,in}$ is a measure of the average drive amplitude during initialization, and not the maximum drive amplitude. At the end of the initialization, $\Lambda_{1,in}$ must be ramped up to match $\Lambda_1$, but $\Lambda_1$ will be approached from a smaller value and therefore the maximum value of $\Lambda_{1,in}$ will remain smaller than $\Lambda_{1}$ during the initialization for this particular choice of $\tau$.

\subsection{Initialization and Pump Errors}
In the previous work on unconditional Fock state generation, the leakage of the blockade state to the higher-populated states due to a small error in the steady-state single-photon drive amplitude has been discussed \cite{lingenfelter1}. However, in an experiment, many other sources of noise are present. One potential limiting experimental factor is the initialization duration ($\tau$). During the initialization, the nonlinearity will apply a squeezing effect to the state. When the initialization time is much smaller than the loss rate of the cavity, this effect can be ignored. However, for large $\kappa$, this requirement would lead to an experimentally infeasible initialization duration. In addition, $\Lambda_1$ for the initialization step scales as $1/\tau$, so one would like to be able to initialize using longer $\tau$ to keep the pump powers minimal. Together, these two factors also make it essential to have quality factors as large as possible. 

\begin{figure*}
\includegraphics[width=\textwidth]{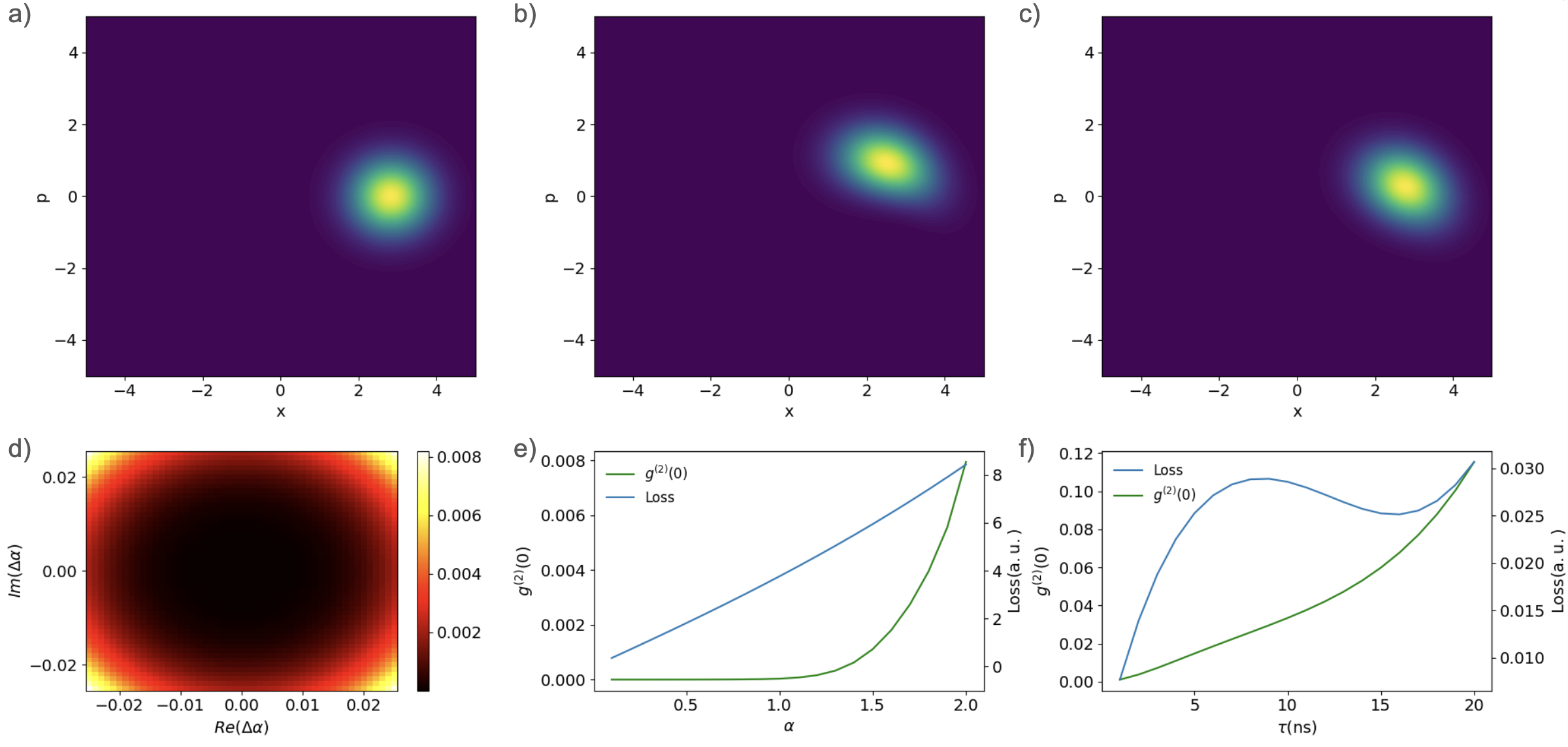}
\caption{\label{fig:alpha error vs g2} Top: Wigner distributions of the state after various initialization conditions. a) Linear cavity, one-photon pump. The state is initialized to a coherent state $\ket{\alpha}$. b) Nonlinear cavity, no two-photon pump. The coherent state is squeezed and shifted slightly. c) Nonlinear cavity, one- and two-photon pumps. The state is moved back to the correct displacement $\alpha$, and the squeezing is less pronounced. Bottom: The individual effect of errors on the final $g^{(2)}(0)$. d) The relationship between $\Delta\alpha$ and $g^{(2)}(0)$ for $\alpha = 2$. e) The error in initialization with an error of $\Delta\alpha = 0.035\alpha$ vs. $\alpha$. f) Under the current protocol, the optimal $g^{(2)}(0)$ given an initialization time $\tau$, assuming $\kappa = 100 \mathrm{\, MHz}$. For e) and f), the blue line represents the loss function that was used for optimization. While there is some mismatch, changing the loss function has not shifted the results significantly. All the subfigures pertain to the standard cavity parameters ($Q = 10^7 $ and $V_{eff} = 0.01 \mu m^3$).}
\end{figure*}

In Fig. \ref{fig:alpha error vs g2}(a)\textendash(c), the Wigner distribution plots demonstrate the squeezing during the initialization and the impact of using a two-photon drive. In \ref{fig:alpha error vs g2}(a), we can see a near-perfect initialization to a coherent state for a linear cavity. In the presence of finite nonlinearity, the displaced state is a squeezed state (Fig. \ref{fig:alpha error vs g2}(b)), and inclusion of a two-photon drive reduces the squeezing (Fig. \ref{fig:alpha error vs g2}(c)). All three plots were obtained using the optimized pumping scheme for each case.  

We also study the impact of initialization errors as an error in the displacement parameter $\alpha$. To quantify the impact of initialization errors on blockade, we study the changes in the zero-delay second-order correlation function, $g^{(2)}(0)$. For this, we only consider the effect on $g^{(2)}(0)$ in the short time ($t \approx 1/\kappa$). The heatmap for variation in $g^{(2)}(0)$ is plotted as a function of small fractional error $\delta\alpha$ in Fig. \ref{fig:alpha error vs g2}(d) for $\alpha =2$. The $g^{(2)}(0)$ only depends on the amplitude of the $\delta\alpha$ and we can see that even for errors larger than 1\%, the $g^{(2)}(0)$ is less than $10^{-3}$. Although the blockade is somewhat resilient to errors during the short timescale, increasing $\alpha$ does result in a rapid rise in the $g^{(2)}(0)$ for the same fractional error. For $\alpha$ within the range 0 to 2, we show this behavior in Fig. \ref{fig:alpha error vs g2}(e). The fractional error of $3.5$\% is kept constant for all the $\alpha$ values in the plot. So for higher photon generation probability, the accurate initialization is very necessary. In Fig. \ref{fig:alpha error vs g2}(f), we show the effect of changing the $\tau$ on $g^{(2)}(0)$ and the amplitude of the loss function. A nonlinear growth in the loss function can be seen. The $g^{(2)}(0)$ rises rapidly for shorter $\tau$, but doesn't show a monotonic behavior for larger values of $\tau$.

\begin{figure*}
\includegraphics[width=\textwidth]{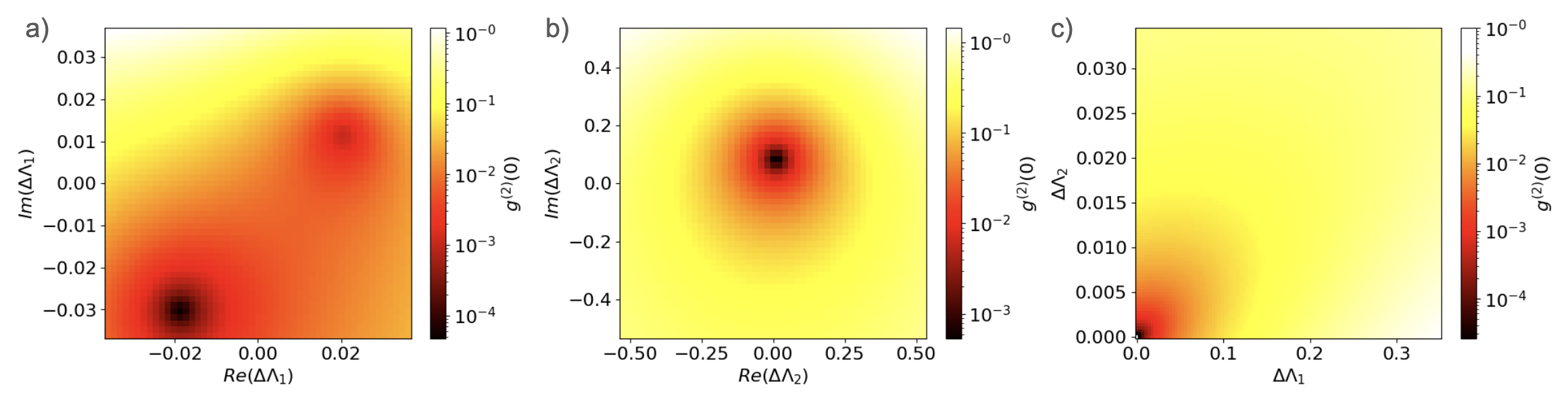}
\caption{\label{fig:combined_errs} Heatmaps of the relationship between errors in $\Lambda_1$ and $\Lambda_2$ with the resulting $g^{(2)}(0)$. All errors are relative to the magnitude of the drives. Parameters:  a) $\Lambda_1$ error during initialization. Multiple minima are present. b) $\Lambda_2$ error during initialization. c) $\Lambda_1$ and $\Lambda_2$ error during evolution in the displaced frame.}
\end{figure*}

While errors in $\alpha$ provide us some insight into the propagation of errors in the protocol, they are not necessarily the types of errors that would occur in the experiment, because the initialized state could be a non-coherent state, as we can see in Fig. \ref{fig:alpha error vs g2} (c). Hence, we also simulate the effect of the errors in pump amplitudes. From Eq. (\ref{lambda_1 linear}), an error in $\Lambda_1$ during intialization translates roughly to a linear error in $\alpha$ in the absence of nonlinearity. However, the presence of nonlinearity shifts the ideal $\Lambda_1$ requirement, and it also causes the solution space to be non-convex. Both of these can be seen in Fig. \ref{fig:combined_errs} (a). Here, $\Delta \Lambda_1$ is the fractional error of $\Lambda_1$.

Similarly, we study the effect of error in the two-photon drive (Fig. \ref{fig:combined_errs}(b)) where we plot the $g^{(2)}(0)$ for a range of fractional error ($\Delta \Lambda_2$) in $\Lambda_2$. Such an error can be caused by individual errors in the single-photon pumps used to generate the two-photon drive, and/or by two-photon loss in the medium. We find that the experiment is more tolerant to two-photon drive error in the initialization phase, and even with around $20\%$ error, the $g^{(2)}(0)$ stays below $0.1$. Interestingly, the lowest $g^{(2)}(0)$ values are not centered at zero $\Lambda_2$ error. It shows us that the state initialized by minimizing the loss function is not necessarily the state that leads to the strongest photon blockade. It is likely the case because the intialized state is not a perfect displaced coherent state, as we can see in Fig. \ref{fig:alpha error vs g2}(c). 

The combined effect of errors in $\Lambda_1$ and $\Lambda_2$ during the evolution in the displaced frame on the final $g^{(2)}(0)$ is shown in Fig. \ref{fig:combined_errs}(c). Unlike the case of initialization errors, we see that the fractional errors in $\Lambda_2$ become more crucial in this phase of the protocol. Furthermore, the $g^{(2)}(0)$ is minimized for zero pump errors.  

\section{Discussion}
We have shown that a triply-resonant silicon photonic crystal cavity could be a good candidate to achieve single-photon blockade via the unconditional Fock state generation mechanism. A big challenge is the simultaneous realization of the high quality factor and the mode volume used in this study. There are examples in literature that have come close to such realization with silicon-based singly-resonant nanobeams \cite{snijders2018observation}. Inverse design might prove vital to achieve it for a triply resonant cavity. 

One avenue of improvement for this protocol is further optimization of the pumping scheme for initialization. It would be both an interesting theoretical as well as experimental problem to include higher-order polynomial ramping instead of just the linear ramping, to minimize the displacement errors. An essential next step would be the more accurate modelling of the two-photon drive by considering other nonlinear processes, such as cross-phase modulation, into account. Such a model should also aid in minimizing optical bistability in degenerate optical parametric oscillations, which could significantly limit this implementation.

\begin{acknowledgments}
The research is supported by the National Science Foundation (NSF) Science and Technology Center (STC) for Integration of Modern Optoelectronic Materials on Demand (IMOD) under Cooperative Agreement No. DMR-2019444. We acknowledge Dr. Avik Dutt for discussion on nonlinear optical effects. M.P. and M.D.S. would like to acknowledge NSF-REU programme under NSF grant No. - 2243362. M.D.S. thanks Dr. David Kagan for valuable discussions and broader scientific guidance.

\end{acknowledgments}

\appendix

\bibliography{mainbib}

\appendix

\section{Derivation of the expression for the input power in Equation (11)}

The cavity input$\mbox{--}$output relations, for mode 1 with a single-photon pump is \cite{Quantum2007}
\begin{equation}
\frac{d\langle\hat{a}\rangle}{dt} = \Delta \hat{a} - \frac{\kappa}{2} \hat{a} +  \sqrt{\kappa}\hat{a}_{in}(t)    
\end{equation}
where $\Delta$ is the cavity mode$\mbox{--}$pump detuning and $\hat{a}_{in}(t)$ is the input field operator which is related to the one-photon drive amplitude as $\sqrt{\kappa}\langle\hat{a}_{in}(t)\rangle = \Lambda_1(t)$. Furthermore, we have assumed that the cavity has only one loss channel with loss rate $\kappa$. 
Also, $\hat{a}_{in}(t)$ is related to the input power ($P_1$) by following relation
\begin{equation}
|\langle \hat{a}_{in}(t) \rangle|^2 = \frac{P_1}{\hbar \omega} 
\end{equation}

Therefore, we get
\begin{equation}
P_1 = \frac{\hbar \omega |\Lambda_1|^2}{\kappa}  
\label{power_1}
\end{equation}

\section{Initialization with a single pump without nonlinearity}
The one-photon pumping Hamiltonian, assuming no dissipation and detuning, has the following form
\begin{equation}
\hat{H_1} = \Lambda_1\hat{a} + \Lambda_1^{\ast}\hat{a}^{\dagger}    
\end{equation}
which gives the Heisenberg's equation of motion for the annihilation operator
\begin{equation}
\frac{d\langle\hat{a}\rangle}{dt} = -i\Lambda_1  
\end{equation}
Starting with a vacuum state at time $t=0$, at time $\tau$ we get
\begin{gather}
\langle\hat{a}\rangle(\tau) = \alpha = -i\Lambda_1\tau + \langle\hat{a}\rangle(0) = -i\Lambda_1\tau \nonumber \\
\Rightarrow \Lambda_1 = \frac{i\alpha}{\tau} \label{eq:final}
\end{gather}

\section{Derivation of the expression for the coupling constant in Equation (6)}
\label{beta}
The Kerr interaction term in the Hamiltonian has the form
\begin{equation}
   \hat{H}_{Kerr} = \int\frac{1}{2}\hat{E}(\vec{r}).\hat{P}^{(3)}(\vec{r})d^3\vec{r}
\end{equation}
where $\hat{E}$ is the total electric field operator and $\hat{P}^{(3)}$ is the third-order polarization, with the following forms
\begin{equation}
\hat{E} = \displaystyle\sum_{i=1}^3\hat{E}_{i}
\end{equation}
\begin{equation}
\hat{P}^{(3)}_i = \epsilon_o\displaystyle\sum_{jkl}\chi_{ijkl}^{(3)} \hat{E}_j\hat{E}_k\hat{E}_l
\end{equation}
where, subscript $i$ denotes the $i$th component of the total electric field $\hat{E}$ and the spatial dependence is kept implicit for brevity. In what follows, we use Roman letters for the vector component index and Greek letters for the mode index. Then, these components could be further expanded as the sum of the electric field components of individual modes    
\begin{equation}
\hat{E}_i = \displaystyle\sum_{\nu=1}^3\hat{E}_{\nu,i}
\end{equation}
where $\hat{E}_{\nu,i}$ is the $i^{th}$ component of the electric field of the $\nu^{th}$ mode. 

Furthermore, the electric field operator is
\begin{equation}
\hat{E}_{\nu} = \displaystyle\sum_{i=1}^3\sqrt{\frac{\omega_{{\nu}c}}{2\bar{\epsilon}_{\nu} V_{\nu}}}\phi_{\nu,i} (\vec{r}) \hat{a}_{\nu} + h.c.
\end{equation}

where, $\bar{\epsilon}_{\nu}$ is the permittivity at a reference location (the location of the field maxima) and $V_{\nu}$ is the mode volume of the $\mu$th mode defined as 
\begin{equation}
    V_{\nu} = \int d^3 \vec{r'} \frac{{\epsilon}_{\nu}(\vec{r'})}{\bar{\epsilon}_{\nu}}|\phi_{\nu} (\vec{r'})|^2 
\end{equation}
with $\bar{\epsilon}_{\nu} = {\epsilon_{\nu}}(\vec{r}_{ref})$, where ${\epsilon}_{\nu}(\vec{r'})$ is the permittivity at location $\vec{r'}$ and $|\phi_{\nu} (\vec{r}_{ref})|^2 = 1$.

Then, in the Hamiltonian, the degenerate four-wave mixing process for two-photon drive has contribution from the terms of the form $\phi_1^{*2}(\vec{r})\phi_2(\vec{r})\phi_3(\vec{r})$ and there are three such terms. Comparing the resulting expression for $\hat{H}_{Kerr}$ with the two-photon drive Hamiltonian in equation \ref{H_2ph}, we get

\begin{equation}
    \beta = \frac{3\epsilon_0}{8}\sqrt{\frac{\omega_{1c}^2 \omega_{2c} \omega_{3c}}{ V_1 ^2 V_2 V_3 \bar{\epsilon}_{1}^2 \bar{\epsilon}_{2} \bar{\epsilon}_{3}}}\int d^3\vec{r} \displaystyle\sum_{ijkl}\chi_{ijkl}^{(3)}\phi^*_{1,i}\phi^*_{1,j}\phi_{2,k}\phi_{3,l}
\end{equation}
The spatial dependence is once again kept implicit. We make further approximation of identically polarized modes to remove the summation and use the relevant component $\chi^{(3)}$ of the tensor $\chi^{(3)}_{ijkl}$. This gives us the final expression for the coupling constant

\begin{equation}
    \beta = \frac{3\epsilon_0}{8}\sqrt{\frac{\omega_{1c}^2 \omega_{2c} \omega_{3c}}{ V_1 ^2 V_2 V_3 \bar{\epsilon}_{1}^2 \bar{\epsilon}_{2} \bar{\epsilon}_{3}}}\int d^3\vec{r} \chi^{(3)}(\vec{r}) \phi_1^{*2}(\vec{r})\phi_2(\vec{r})\phi_3(\vec{r})
    \label{beta_eq}
\end{equation}

\end{document}